\documentstyle[multicol,epsf,aps]{revtex}
\begin{document}
\title{Genetic Correlations in Mutation Processes}
\author{E.~Ben-Naim and A.~S.~Lapedes}
\address{Theoretical Division and Center for Nonlinear Studies, Los Alamos 
National Laboratory, Los Alamos NM 87545}
\maketitle
\begin{abstract}
We study the role of phylogenetic trees on correlations in mutation
processes.  Generally, correlations decay exponentially with the
generation number.  We find that two distinct regimes of behavior
exist. For mutation rates smaller than a critical rate, the underlying
tree morphology is almost irrelevant, while mutation rates higher than
this critical rate lead to strong tree-dependent correlations. We show
analytically that identical critical behavior underlies all multiple
point correlations. This behavior generally characterizes branching
processes undergoing mutation.

\end{abstract}

\begin{multicols}{2} 
\section{Introduction}
Biological evolution is influenced by a number of processes including
population growth, mutation, extinction, and interaction with the
environment, to name a few \cite{Kimura}. Genetic sequences are
strongly affected by such processes and thus provide an important clue
to their nature.  The ongoing effort of reconstructing evolution
histories given the incomplete set of mapped sequences constitutes
much of our current understanding of biological evolution.

However, this challenge is extraordinary as it involves an inverse
problem with an enormous number of degrees of freedom. Statistical
methods such as maximum likelihood techniques coupled with simplifying
assumptions on the nature of the evolution process are typically used
to infer the structure of the underlying evolutionary tree, {\it
i.e.}, the phylogeny \cite{Waterman,Durbin,Maxent,Kishino}.

Genetic sequences such as RNA/DNA or amino acid sequences can be seen
as words with letters taken from an alphabet of 4 or 20 symbols,
respectively. Generally, there are nontrivial intra-sequence
correlations that influence the evolution of the entire sequence.
Additionally, the structure of the evolutionary tree plays a role in this
process as one generally expects that the closer sequences are on this
tree, the more correlated they are \cite{Altschul}.  In this study, we
are interested in describing the influence of the latter aspect,
namely the phylogeny, on the evolution of sequences. Specifically, we
examine correlations between sequences, thereby complementing related
studies on changes in fluctuations and entropy due to the phylogeny
\cite{Lapedes,Giraud}.  To this end, we consider particularly simple
sequences and focus on a model that mimics the competition between the
fundamental processes of mutation and duplication. 

The rest of this paper is organized as follows.  In Sec.~II, the model
is introduced, and the main result is demonstrated using the pair
correlations.  Correlations of arbitrary order are obtained and 
analyzed asymptotically in Sec.~III.  To examine the range of validity
of the results, generalizations to stochastic tree morphologies and
sequences with larger alphabets are briefly discussed in Sections IV
and V.  Sec.~VI discusses implications for multiple site correlations
in sequences with independently evolving sites. We conclude with a
summary and a discussion in Sec.~VII.

\section{Pair Correlations}

Let us formulate the model first. The sequences are taken to be of
unit length and the corresponding alphabet consists of two letters.
The numeric values $\sigma=\pm 1$ are conveniently assigned to these
letters.  We will focus on binary trees where the number of children
equals two.  This structure is deterministic in that both the number
of children and the generation lifetime are fixed.  Nevertheless, the
results apply qualitatively to stochastic tree morphologies as well.
Finally, the mutation process is implemented as follows: with
probability $1-p$ a child equals his predecessor while with
probability $p$ a mutation occurs, as illustrated in Fig.~1.  The
mutation process is invariant under the transformation
$\sigma\to-\sigma$ and $p\to 1-p$, and we restrict our attention to
the case $0\leq p\leq 1/2$ without loss of generality.

\begin{figure}
\centerline{\epsfxsize=9cm \epsfbox{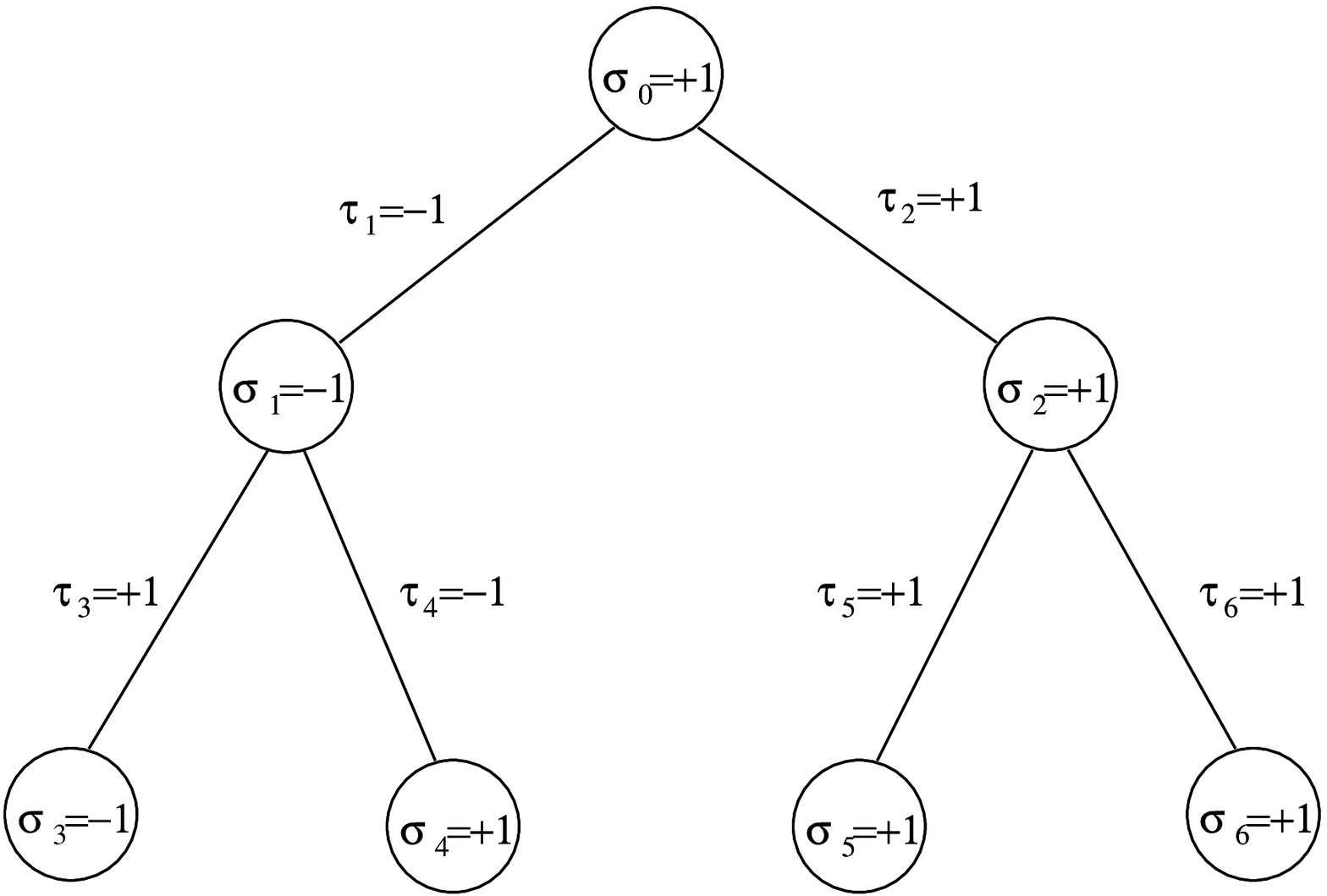}}
\noindent{\small {\bf Fig.1} The mutation process on a two-generation
binary tree. The multiplicative variable $\tau$ indicates whether a
mutation occurred.}
\end{figure}

A natural question is how correlated are the various leafs (or nodes)
of a tree in a given generation (or equivalently, time)? Consider
$G_2(k)$ the average correlation between two nodes at the $k$th
generation
\begin{equation}
\label{pair}
G_2(k)=\langle\langle \sigma_i\sigma_j\rangle\rangle.
\end{equation}
The first average should be taken over all realizations for a fixed
pair of nodes $i \neq j$, while the second average is taken over all
different pairs belonging to the same generation.  For example,
consider this quantity at the second generation (see Fig.~1),
\hbox{$G_2(2)=[\langle\sigma_3\sigma_4\rangle+\langle\sigma_3\sigma_5\rangle+
\langle\sigma_3\sigma_6\rangle]/3$}. One index ($i=3$) may be fixed since
all nodes in a given generation are equivalent.

To evaluate averages, it is useful to assign a multiplicative
random variable $\tau_i=\pm 1$ to every branch of the tree such that
$\sigma_i=\sigma_j\tau_i$ with $j$ the predecessor of $i$. One has
$\tau_i=1$ ($-1$) with probability $1-p$ ($p$), and consequently,
\begin{equation}
\label{tauav}
\langle\tau\rangle\equiv\langle \tau_i\rangle=1-2p. 
\end{equation}

Pair correlations are readily calculated using the $\tau$ variables:
writing $\sigma_3=\sigma_0 \tau_1\tau_3$ and similarly for $\sigma_4$
gives $\langle\sigma_3\sigma_4\rangle=\langle \sigma_0\tau_1\tau_3
\sigma_0\tau_1\tau_4\rangle=\langle\sigma_0^2\tau_1^2\tau_3\tau_4\rangle$.
Since \hbox{$\sigma_i^2=\tau_i^2=1$}, this correlation simplifies,
$\langle\sigma_3\sigma_4\rangle=\langle\tau_3\tau_4\rangle$.
Furthermore, mutation processes on different branches are independent
and consequently $\langle\tau_i\tau_j\rangle= \langle
\tau_i\rangle\langle\tau_j\rangle$ when $i\neq j$. Thus, $\langle
\sigma_3\sigma_4 \rangle=\langle \tau\rangle^2$ and similarly
\hbox{$\langle\sigma_3\sigma_5\rangle=\langle\sigma_3\sigma_6\rangle=\langle\tau\rangle^4$}. The
overall picture becomes clear: when calculating two-point
correlations, the path to the tree root is traced for each node. As
$\tau^2=1$, doubly counted branches cancel. Only branches that trace
the path to the first common ancestor are relevant.  In other
words,  
\begin{equation}
\label{twopoint}
\langle\sigma_i\sigma_j\rangle=\langle\tau\rangle^{d_{i,j}}
\end{equation}
with $d_{i,j}$  the ``genetic distance'' between two points, the
minimal number of branches that connect two nodes.  Indeed, at the
second generation $d_{3,4}=2$, $d_{3,5}=d_{3,6}=4$ and consequently
\hbox{$G_2(2)=(\alpha^2+2\alpha^4)/3$} with the shorthand notation
\hbox{$\alpha=\langle\tau\rangle=1-2p$}. This generalizes into a geometric
series
\hbox{$G_2(k)=(\alpha^2+2\alpha^4+\cdots+2^{k-1}\alpha^{2k})/(2^k-1)$}. Evaluating
this sum gives the pair correlation 
\begin{equation}
\label{g2ka}
G_2(k)={\alpha^2\over 2\alpha^2-1}{(2\alpha^2)^k-1\over 2^k-1}.
\end{equation}
Interestingly, pair correlations are not affected by the initial
state, {\it i.e.}, the value of the tree root.

For sufficiently large generation numbers, the leading order of the
pair correlation decays exponentially with the generation
number. However, different constants characterize this decay,
depending on the mutation probability
\begin{equation}
\label{g2kb}
G_2(k)\simeq\cases{{\alpha^2\over 2\alpha^2-1}\,\alpha^{2k}&$p<p_c$;\cr
{\alpha^2\over 1-2\alpha^2}\,2^{-k}&$p>p_c$.}
\end{equation}
As seen from Eq.~(\ref{g2ka}), the transition between the two different 
behaviors occurs when $2\alpha^2=1$ or alternatively at the following 
mutation probability  
\begin{equation}
\label{pc}
p_c={1\over 2}\left(1-{1\over\sqrt{2}}\right). 
\end{equation}
Although in general correlations decay exponentially $G_2(k)\sim
\beta^{2k}$, the decay constant $\beta$ exhibits two distinct
behaviors which depend on the mutation probability $\alpha$. When the
mutation probability is smaller than the critical one $p<p_c$ then 
$\beta=\alpha$ while in the complementary case $\beta=1/\sqrt{2}$.

As a reference, it is useful to consider the decay of the
average node value $G_1(k)=\langle\sigma\rangle$.  At the $k$th
generation, the path to each node involves $k$ branches and thus,
$G_1(k)= G_1(0)\alpha^k$ with $G_1(0)=\langle
\sigma_0\rangle$.  Writing $G_1(k)\sim \beta^k$ then $\beta=\alpha$ for all
mutation probabilities, in contrast with the asymptotic behavior of
$G_2(k)$. Below the critical mutation rate, $G_2(k)\propto
[G_1(k)/G_1(0)]^2$, indicating that knowledge of the one-point average
suffices to characterize correlations. 

\begin{figure}
\centerline{\epsfxsize=9cm \epsfbox{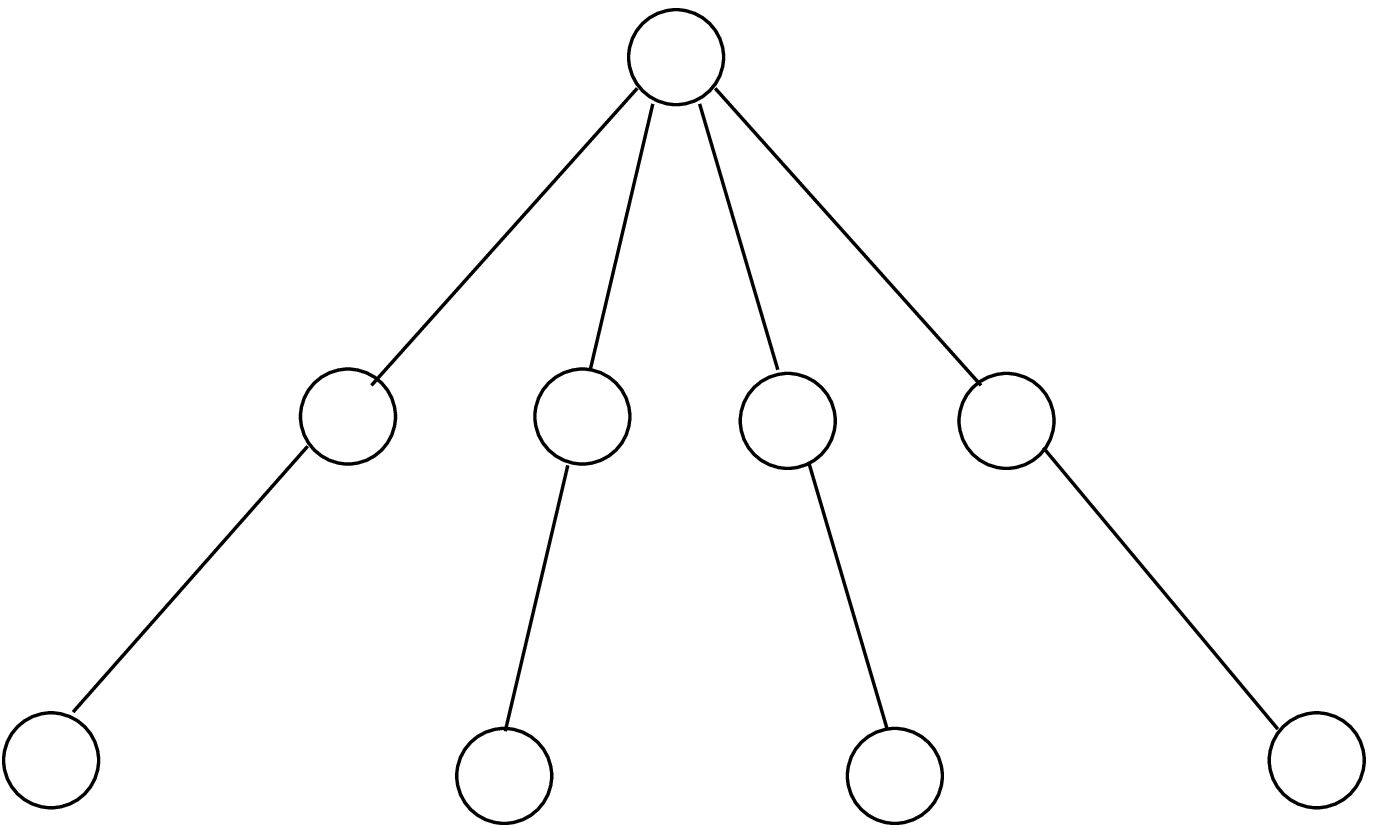}}
\noindent{\small {\bf Fig.~2} The trivial ``star'' Phylogeny. The path
connecting two nodes always contains the tree root.} 
\end{figure}

In fact, the above behavior can be attributed to the tree morphology.  To
see that, it is useful to consider a structureless morphology where
the only ancestor shared by two nodes is the tree root itself (see
Fig.~2).  Using the notation $G^*$ to denote correlations on this
``star'' morphology, we see that the average remains unchanged
$G_1(k)=G^*_1(k)=G_1^*(0)\alpha^k$.  The star morphology is trivial in
that all genetic distances are equal: $d_{i,j}=2k$ when $i\neq
j$. Thus, pair correlations are immediately obtained from the average
$G^*_2(k)=[G^*_1(k)/G^*_1(0)]^2=\alpha^{2k}$.  As branches in the star
morphology do not interact, no correlations develop.

In contrast, nontrivial phylogenies do induce correlations. Indeed,
$G_2(k)>G_2^*(k)$ when $p>0$.  Interestingly, when $p<p_c$, merely the
asymptotic prefactor  $\alpha^2/(2\alpha^2-1)>1$ in
Eq.~(\ref{g2kb}) is enhanced and $G_2(k)\propto G_2^*(k)$. As the
critical point is approached, this constant diverges thereby signaling
the transition into a second regime. When $p>p_c$, the decay constant
itself is enhanced and the ratio $G_2(k)/G_2^*(k)$ grows
exponentially. The mutation probability affects only the asymptotic
prefactor, and the decay constant $\beta=1/\sqrt{2}$ is determined by
the tree morphology. We conclude that the nontrivial phylogeny
generates significant correlations for larger than critical mutation
probabilities.

This behavior can be understood and partially rederived using a 
heuristic argument. Genetically close nodes are highly
correlated, while distant pairs are weakly correlated, as indicated by
Eq.~(\ref{twopoint}). On the other hand, distant pairs are
more numerous. Both effects are magnified exponentially for large
generation numbers, and their competition results in a critical point.
Different mechanisms dominate on different sides of this point.
Specifically, the number of minimal genetic distance pairs ($d=2$) is
$2^{k-1}$, while the number of maximal distance pairs ($d=2k$) is
$2^{2(k-1)}$. The rule (\ref{twopoint}) gives the relative
contributions of these two terms to the overall two-point correlation:
$2^{k-1}\alpha^2$ versus $2^{2(k-1)}\alpha^{2k}$. These are simply the
first and last terms in the geometric series that led to
Eq.~(\ref{g2ka}). Comparing these two terms in the limit $k\to\infty$
correctly reproduces the most relevant aspects, {\it i.e.}, the location of
the critical point (\ref{pc}) and the decay constants of
Eq.~(\ref{g2kb}).  We conclude that competition between the
multiplicity and the degree of correlation of close and distant nodes
underlies the transition.

\section{Higher Order Correlations}

The above analysis gives useful intuition for the overall qualitative
behavior. Yet, it can be generalized into a more complete treatment
that addresses correlations of arbitrary order.  This set of
quantities is helpful in determining the extent to which this picture
applies, and in particular, whether the transition is actually a phase
transition.

Multiple point correlations obey a rule similar to
Eq.~(\ref{twopoint}).  For example, consider the four-node average
$\langle \sigma_3\sigma_4\sigma_5\sigma_6\rangle$ in Fig.~1. Using the
$\tau$ variables, we rewrite
$\langle\sigma_3\sigma_4\sigma_5\sigma_6\rangle=\langle
\sigma_0^4\tau_1^2\tau_2^2\tau_3\tau_4\tau_5\tau_6\rangle$, and since
$\sigma^2=\tau^2=1$ we get
$\langle\sigma_3\sigma_4\sigma_5\sigma_6\rangle=
\langle\tau_3\tau_4\tau_5\tau_6\rangle=\langle\tau \rangle^4$ or
$\langle\sigma_3\sigma_4\sigma_5\sigma_6\rangle=
\langle\sigma_3\sigma_4\rangle\langle\sigma_5\sigma_6\rangle$.  The
four point average equals a product of two-point averages with the
indices chosen as to minimize the total number of branches. This 
can also be seen by tracing the path of each node to the tree root and
canceling doubly counted branches. Thus, Eq.~(\ref{twopoint})
generalizes as follows: 
\begin{equation}
\label{fourpoint}
\langle\sigma_i\sigma_j\sigma_k\sigma_l\rangle=
\langle\tau\rangle^{d_{i,j,k,l}},
\end{equation}
with the four-point genetic distance 
\begin{equation}
\label{dijkl}
d_{i,j,k,l}={\rm min}\{d_{i,j}+d_{k,l},d_{i,k}+d_{j,l},d_{i,l}+d_{j,k}\}.
\end{equation}
Similarly, the law for arbitrary order averages is
$\langle\tau\rangle$ raised to a power equal to the $n$-point genetic
distance. Such distance is obtained by considering all possible
decompositions into pairs of nodes. The genetic distance is the
minimal sum of the corresponding pair distances. Averages over an
odd number of nodes can be obtained by adding a ``pseudo'' node at the
root of the tree and using the convention $d_{i,{\rm root}}=k$ when
$i$ belongs to the $k$th generation. The average $\langle
\sigma_0\rangle$ is generated by the root and this factor multiplies
all odd order correlation.  Since even order correlations are
independent of the root value, and odd correlations are simply
proportional to $\langle \sigma_0\rangle$, we set $\langle
\sigma_0\rangle=1$ in what follows without loss of generality.

The average $n$-point correlation is defined as follows
\begin{equation}
\label{gnk}
G_n(k)=\langle\langle\sigma_{i_1}\sigma_{i_2}\cdots\sigma_{i_n}\rangle\rangle,
\end{equation}
where the averages are taken over all realizations and over all
possible choices of $n$ distinct nodes at the $k$th generation.  For
the trivial star phylogeny, the $n$-point genetic distance is constant
and equals a product of the correlation order and the generation
number, $d=nk$. Consequently, all averages are trivial as knowledge of
the one-point average immediately gives all higher-order averages,
\hbox{$G^*_n(k)=[G^*_1(k)]^n$}, or explicitly
\begin{equation}
\label{gnkstar}
G^*_n(k)=\alpha^{nk}.
\end{equation} 

When the tree morphology is nontrivial, the minimal-sum rules
(\ref{fourpoint})-(\ref{dijkl}) imply that such factorization no
longer holds.  For binary trees, it is possible to obtain these
correlations recursively. Let us assign the indices $1,2,\ldots,2^k$
to the $k$th generation nodes and order them as follows $1\leq
i_1<i_2<\cdots<i_n\leq 2^k$.  As the average over the realizations is
performed first, the average correlation requires a summation over all
possible choices of nodes
\begin{equation}
\label{fnkdef}
F_n(k)=\sum_{1\leq i_1<i_2<\cdots<i_n\leq 2^k}
\langle\sigma_{i_1}\sigma_{i_2}\cdots\sigma_{i_n} \rangle.
\end{equation}
Proper normalization gives the  $n$-node correlation
\begin{equation}
\label{gnkfnk}
G_n(k)=F_n(k)\Big/ {2^k \choose n}.   
\end{equation}

Consider a group of $n$ nodes taken from the $k$th generation. They
all share the tree root as a common ancestor. The two first generation
nodes naturally divide this group into two independently evolving
subgroups. This partitioning procedure allows a recursive calculation
of the correlations. Formally, a given choice of nodes \hbox{$1\leq
i_1<i_2<\cdots<i_n\leq 2^k$} is partitioned into two subgroups as
follows \hbox{$1\leq i_1<\cdots<i_m\leq 2^{k-1}$} and
\hbox{$2^{k-1}+1\leq i_{m+1}<\cdots<i_n\leq 2^{k-1}+2^{k-1}$}. These
subgroups involve different $\tau$ variables, so their correlations
factorize
\begin{equation}
\label{factorize}
\langle\sigma_{i_1}\cdots\sigma_{i_n}\rangle
\propto \langle\sigma_{i_1}\cdots\sigma_{i_m}\rangle
\langle\sigma_{i_{m+1}}\cdots\sigma_{i_n}\rangle.
\end{equation}
The proportionality constant depends upon the parity of $m$ and $n-m$.
Even correlations are independent of the tree root, while odd
correlations are proportional to the average value of the tree
root. This extends to sub-trees as well, and since $\sigma_0=1$, the
average value of the root of both sub-trees is $\langle\tau\rangle$.
This factor accompanies all odd correlations. Substituting
Eq.~(\ref{factorize}) into Eq.~(\ref{fnkdef}) shows that the summation
factorizes as well.  Using \hbox{$F_m(k-1)=\sum_{1\leq i_1<\cdots<
i_m\leq 2^{k-1}} \langle\sigma_{i_1}\cdots\sigma_{i_m}\rangle$}
reduces the problem to two sub-trees that are one generation shorter,
and a recursion relation for $F_n(k)$ emerges
\begin{equation}
\label{fnk}
F_n(k)=\sum_{m=0}^n F_m(k-1)B_mF_{n-m}(k-1)B_{n-m}, 
\end{equation}
with the boundary conditions $F_n(0)=\delta_{n,0}+\delta_{n,1}$. The
summation corresponds to the $n+1$ possible partitions of a group
of $n$ nodes into two subgroups. The weight of the odd correlations is
accounted for by $B_n$
\begin{equation}
\label{bn}
B_n=\cases{1&$n=2r$;\cr \langle \tau\rangle&$n=2r+1$.\cr}
\end{equation}
Using the definition (\ref{fnkdef}), the sums $F_n(k)$ vanish whenever
$n>2^k$. This behavior emerges from the recursion relations as well.
Additionally, one can check that the sums are properly normalized in
the no mutation case ($\alpha=1$), $F_n(k)={2^k \choose n}$ when
$n\le 2^k$.

For sufficiently small $n$, it is possible to evaluate the 
sums explicitly using Eqs.~(\ref{fnk}). The average correlations are 
then found using Eq.~(\ref{gnkfnk})
\begin{eqnarray}
\label{g123}
G_0(k)&=&1,\nonumber\\ 
G_1(k)&=&\alpha^k,\nonumber\\ 
G_2(k)&=&{\alpha^2\over 2\alpha^2-1}{(2\alpha^2)^k-1\over 2^k-1},\\ 
G_3(k)&=&{3\alpha^{k+2}\over 2\alpha^2-1}
{{(4\alpha^2)^k-(4\alpha^2)\over 4\alpha^2-1}-(2^k-2 )\over 
(2^k-1)(2^k-2)}.\nonumber
\end{eqnarray}
Indeed, these quantities agree with the previous results for $n=1$, $2$
and equal unity when $p=0$. We see that correlations involve a sum
of exponentials. Furthermore, it appears that the condition
$2\alpha^2=1$ still separates two different regimes of
behaviors. However, calculating higher correlations explicitly is not
feasible as the expressions are involved for large $n$.  Instead, we
perform an asymptotic analysis that more clearly exposes the leading
large generation number behavior.

Let us consider first the regime $p<p_c$ or equivalently
$2\alpha^2>1$. From Eq.~(\ref{g123}), we see that the leading large
$k$ behavior of the average correlation satisfies $G_n(k)\sim
\alpha^{nk}$ for $n=0,1,2$, and $3$. We will show below that this behavior
extends to higher order correlations, {\it i.e.},
\begin{equation}
\label{gnka}
G_n(k)\simeq g_n \alpha^{nk}.
\end{equation}
In other words, the following limit $\alpha=\lim_{k\to\infty}
[G_n(k)]^{1/nk}$ exists and is independent of $n$. As correlations are
larger when the phylogeny is nontrivial, one expects that $G_n(k)\ge
G_n^*(k)$ or in terms of the prefactors, $g_n\ge g_n^*=1$.  Combining
Eq.~(\ref{gnkfnk}) with the leading behavior of the combinatorial
normalization constant ${2^k\choose n}\sim 2^{nk}/n!$ gives the
the asymptotic behavior of the sums
\begin{equation}
\label{fnka}
F_n(k)=f_n(2\alpha)^{nk},\quad {\rm with}\quad f_n={g_n\over n!}.  
\end{equation}

Substituting Eq.~(\ref{fnka}) into the recursion relation
Eq.~(\ref{fnk}) eliminates the dependence on the generation number $k$,
and a recursion relation for coefficients $f_n$ is found
\begin{equation}
\label{fn}
f_n (2\alpha)^n=\sum_{m=0}^r f_mB_mf_{n-m}B_{n-m}, 
\end{equation}
with $B_n$ of Eq.~(\ref{bn}). These recursion relations are consistent
with the conditions $f_0=f_1=1$. The case $n=2$ reproduces the
coefficient \hbox{$f_2=\alpha^2/[(2\alpha)^2-2]$}.  The divergence at
$2\alpha^2=1$ indicates that the ansatz (\ref{gnka}) breaks down at
the critical point.  To show that the ansatz holds in the entire range
$0\leq p< p_c$, one has to show that the coefficients $f_n$ are
positive and finite for all $n$. Rewriting the recursion (\ref{fn})
explicitly \hbox{$f_n[(2\alpha)^n-2B_n]=\sum_{m=1}^{r-1}
f_mB_mf_{n-m}B_{n-m}$} allows us to prove this.  Since $f_0=1>0$, then
to complete a proof by induction one needs to show that a positive
$f_{n-1}$ implies a positive $f_{n}$. The right hand side of the
recursion is clearly positive and thus the positivity of $f_n$ hinges
on the positivity of the term $(2\alpha)^n-2B_n$. When $2\alpha^2>1$,
then $\alpha>1/\sqrt{2}$ and certainly $2\alpha>1$.  Combining this
with the inequality $(2\alpha)^2>2>2B_n$ shows that
$(2\alpha)^n-2B_n>0$ when $n\geq 2$. Hence $f_n$ is positive and
finite for all $n$, which validates the ansatz (\ref{gnka}) in the
regime $p<p_c$.

In principle, the coefficients can be found by introducing the
generating functions
\begin{equation}
\label{az}
f(z)=\sum_n f_n z^n.
\end{equation}
Multiplying Eq.~(\ref{fn}) by $z^n$ and summing over $n$ yields 
the following equation for the generating functions
\begin{equation}
\label{a2az}
f(2\alpha z)=\left[{f(z)+f(-z)\over 2}+\alpha{f(z)-f(-z)\over 2}\right]^2.
\end{equation}
This equation reflects the structure of the recursion relations.  A
factor $\alpha$ is generated by each odd-index coefficient and as a
results, the odd part of the generating functions
$[f(z)-f(-z)]/2=f_1z+f_3z^3+\cdots$ is multiplied by
$\alpha$. Although a general solution of this equation appears rather
difficult, it is still possible to obtain results in the limiting
cases. It is useful to check that when $\alpha=1$, the above equation
reads $f(2z)=f^2(z)$ which together with the boundary conditions
$f_0=f_1=1$ gives $f(z)=\exp(z)$ or $f_n={1\over n!}$. As $g_n\to 1$,
the trivial correlations are recovered, $G_n\to G_n^*$ indicating that
role played by the tree morphology diminishes in the no mutation
limit.

In the  limit $p\to p_c^-$ it is possible to extract
the leading behavior of the asymptotic prefactors. Here, it is sufficient
to keep only the highest powers of the diverging term
$1/(2\alpha^2-1)$.  The calculation in this case is identical to the
one detailed below for the case $p>p_c$ and we simply quote the
results
\begin{equation}
\label{sub}
G_n(k)\to\cases 
{{2r!\over r!}\left[{\alpha^2\over 2(2\alpha^2-1)}\right]^r \alpha^{nk}&$n=2r$;\cr
{(2r+1)!\over r!}\left[{\alpha^2\over 2(2\alpha^2-1)}\right]^r \alpha^{nk}&$n=2r+1$.\cr}  
\end{equation}
In this limit, the odd order correlations simply follow from their
even counterparts and for example  \hbox{$f_{2r+1}=f_{2r}$}.

In the complementary case $p>p_c$, it proves useful to rewrite the
recursion relations (\ref{fn}) for the even and odd correlations
separately
\begin{eqnarray}
\label{reo}
F_{2r}(k)&=&\sum_{s=0}^r F_{2s}(k-1)F_{2r-2s}(k-1)\nonumber\\
         &+&\alpha^2\sum_{s=0}^{r-1} F_{2s+1}(k-1)F_{2r-2s-1}(k-1)\\
F_{2r+1}(k)&=&2\alpha\sum_{s=0}^r F_{2s}(k-1)F_{2r-2s+1}(k-1).\nonumber
\end{eqnarray}
The leading asymptotic behavior of Eq.~(\ref{g123}) implies
$F_0(k)=f_0$, $F_1(k)\simeq f_0 (2\alpha)^k$, $F_2(k)\simeq f_2 2^k$,
and $F_3(k)\simeq f_2 2^k (2\alpha)^k$ with $f_0=1$ and
$f_2=\alpha^2/[2-(2\alpha)^2]$. Let us assume that this even-odd
pattern is general 
\begin{eqnarray}
\label{feo}
F_{2r}(k)&=&f_{2r}2^{rk},\\
F_{2r+1}(k)&=&f_{2r}2^{rk}(2\alpha)^k.\nonumber
\end{eqnarray}
Substituting this ansatz into Eq.~(\ref{reo}) shows that the second
summation in the recursion for the even correlations is negligible
asymptotically. Both equations reduce to
\begin{equation}
f_{2r}2^r=\sum_{s=0}^r f_{2s}f_{2r-2s},
\end{equation}
and therefore the pattern (\ref{feo}) holds when $p>p_c$. It is
seen that odd correlators are enslaved to the even ones.

To obtain the coefficients, we introduce the generating functions
$f(z)=\sum_r f_{2r}z^{2r}$ which satisfies $f(0)=1$, $f'(0)=0$ and
$f''(0)=f_2=\alpha^2/[2(1-2\alpha^2)]$.  The recursion relation
translates into the following equation for $f(z)$
\begin{equation}
f\left(\sqrt{2}z\right)=[f(z)]^2. 
\end{equation}
Its solution is \hbox{$f(z)=\exp[(\alpha z)^2/2(1-2\alpha^2)]$}.
Thus, \hbox{$f_{2r}={1\over r!}[\,f_2\,]^r$}. From 
Eqs.~(\ref{gnka})-(\ref{fnka}), the leading asymptotic behavior in the
regime $p_c<p<1/2$ is found
\begin{equation}
\label{super}
G_n(k)\simeq\cases{
{2r!\over r!}\left[{\alpha^2\over 2(1-2\alpha^2)}\right]^r 2^{-kr}&$n=2r$;\cr
{(2r+1)!\over r!}\left[{\alpha^2\over 2(1-2\alpha^2)}\right]^r \alpha^k2^{-kr} & $n=2r+1$.\cr}
\end{equation}
Using the Stirling formula $n!\simeq \sqrt{2\pi n}\, n^n e^{-n}$ it is
seen that the coefficients $g_{2r}$ have nontrivial $r$ behavior as
$g_{2r}=g_{2r+1}/(2r+1)\simeq \sqrt{2}[2\alpha^2/(1-2\alpha^2)]^r r^r$.

The even order correlations have identical asymptotic behavior to the
two point correlation: $\lim_{k\to\infty} [G_{2r}(k)]^{1/2rk}={1\over
\sqrt{2}}$ for all $r$. The odd order correlations behave differently,
however, as this limit depends on the correlation order:
$\lim_{k\to\infty} [G_{2r+1}(k)]^{1/(2r+1)k}={1\over
\sqrt{2}}(\sqrt{2}\alpha)^{1/2r+1}$.  Thus, only in the limit
$r\to\infty$ do the even and odd order correlations agree.  However,
this conclusion is misleading since the decay rate of the (properly
normalized) odd order correlations $G_{2r+1}(k)/G_1(k)\sim G_{2r}(k)$
is identical to that of the even order correlations. We conclude that
the decay rate of two-point correlations characterizes the decay of
all higher order correlations.

From Eqs.~(\ref{sub}) and (\ref{super}), we see that the coefficients
diverge according to 
\begin{equation}
f_{2r}=f_{2r+1}\sim |p_c-p|^{-r} 
\end{equation}
as the critical point is approached, $p\to p_c$. Since the
correlations must remain finite, this indicates that the purely
exponential behavior must be modified when $p=p_c$. Indeed, evaluating
Eq.~(\ref{g123}) at $p=p_c$ yields $F_2(k)\simeq f_2 2^k$ and
$F_3(k)=f_2 2^{3k/2}$ with $f_2=k/4$, {\it i.e.}, the even-odd pattern of
Eq.~(\ref{feo}) is  reproduced. Furthermore, the value of $f_2$
shows that the diverging quantity $1/|1-2\alpha^2|$ is simply
replaced by $k$.  This implies that the coefficients become generation
dependent, $f_n\to f_n(k)$. Assuming the pattern Eq.~(\ref{feo}),
substituting it into Eq.~(25), and following the steps that led to
Eq.~(\ref{super}) yields the critical behavior
\begin{equation}
\label{critical}
G_n(k)\simeq\cases{
{2r!\over r!}\left[{k\over 4}\right]^r 2^{-kr}&$n=2r$;\cr
{(2r+1)!\over r!}\left[{k\over 4}\right]^r 2^{-k(r+1/2)} & $n=2r+1$.\cr}
\end{equation}
Generally, the diverging quantity $1/|1-2\alpha^2|$ is replaced
with the finite (but ever growing) quantity $k$.  The algebraic
modification to the leading exponential behavior in
Eq.~(\ref{critical}) is reminiscent of the logarithmic corrections
that typically characterize critical behavior in second order phase
transitions \cite{Stanley}.

\section{Stochastic Tree Morphologies}

The question arises: how general is the behavior described above?  The
binary tree considered was particularly simple as it involved a fixed
number of children and a fixed generation lifetime. Below we show that
relaxing either of these conditions does not affect the nature of the
results.

Let us first consider tree morphologies with a varying number of
children, {\it i.e.}, the trees are generated by a stochastic
branching process where with probability $P_r$ there are $r$
children. This probability sums to unity $\sum_r P_r=1$, and the
average number of children is given by $\langle r\rangle=\sum_r r
P_r$. As a result, the average number of nodes at the $k$th generation
is $\langle r\rangle^k$, indicating that the tree ``survives'' only if
$\langle r \rangle>1$, a classical result of branching processes
theory \cite{Harris}.  The rule (\ref{twopoint}) is independent of the
tree morphology, and therefore, one can repeat the heuristic argument
in Sec.~II. The extreme contributions to the average pair correlations
have the relative weights $\langle r\rangle^{k-1}\alpha^2$ and
$\langle r\rangle^{2(k-1)}\alpha^{2k}$. Comparing these two terms
asymptotically shows that the critical point is a simple
generalization of Eq.~(\ref{pc})
\begin{equation}
\label{pc1}
p_c={1\over 2}\left(1-\sqrt{{1\over\langle r\rangle}}\,\right). 
\end{equation}
The critical mutation rate varies from $0$ to $1/2$ as the average
ancestry size varies between $1$ and $\infty$. This indicates that
correlations are significant over a larger range of mutation rates for
smaller trees. The heuristic argument also gives the decay constant
$\beta$, and the leading asymptotic behavior of Eq.~(\ref{g2kb}) is
generalized by simply replacing $2$ with $\langle r\rangle$. A more
complete treatment of this problem is actually possible and closely
follows Eq.~(\ref{g2ka}).  Again, the ancestry size $\langle r\rangle$
replaces the deterministic value $2$. As both the results and the
overall behavior closely follow the deterministic case, we do
not detail them here.

A second possible generalization is to morphologies with a varying
generation lifetime. Such tree morphologies can be realized by
considering a continuous time variable. Branching is assumed to occur
with a constant rate $\nu$. For such tree morphologies, the number of
nodes $n(t)$ obeys $\dot n(t)=\nu n(t)$ which gives an exponential
growth $n(t)=e^{\nu t}$. Similarly, the mutation process is assumed to
occur with a constant rate $\gamma$. A useful characteristic of this
process is the autocorrelation $A(t)=\langle
\sigma(0)\sigma(t)\rangle$. To evaluate it's evolution, we note that
$A(t+dt)=(1-\gamma dt)A(t)-\gamma dt A(t)$ when $dt\to 0$. Therefore,
$\dot A(t)=-2\gamma A(t)$ and one finds $A(t)=e^{-2\gamma t}$. The
quantities $n(t)$ and $A(t)$ allow calculation of the average pair
correlation.

Let us pick two nodes at time $t$ and denote their values by
$\sigma_i(t)$ and $\sigma_j(t)$, and let the genetic distance between
these two nodes be $\tau$. Using their first common ancestor
$\sigma_{c}(t-\tau)=\sigma_i(t-\tau)=\sigma_j(t-\tau)$ and the
identity $\sigma^2=1$, their correlation can be evaluated as follows
$\langle\sigma_i(t)\sigma_j(t)\rangle=
\langle\sigma_i(t)\sigma_{c}(t-\tau)\sigma_{c}(t-\tau)\sigma_j(t)\rangle=
\langle\sigma_i(t)\sigma_i(t-\tau)\rangle
\langle\sigma_j(t)\sigma_j(t-\tau)\rangle= A^2(\tau)$. Integrating
over all possible genetic distances gives the average pair correlation
\begin{equation}
\label{integ}
G_2(t)={\int_0^t d\tau\, n(\tau) A^2(\tau)\over \int_0^t d\tau\, n(\tau)}.
\end{equation}
The factor $n(\tau)/\int_0^t d\tau\, n(\tau)$ accounts for the
multiplicity of pairs with genetic distance $\tau$. Using
$A(t)=e^{-4\gamma t}$ and $n(\tau)=e^{\nu t}$, the average 
pair correlation is evaluated 
\begin{equation}
\label{stoch}
G_2(t)={\nu\over \nu-4\gamma}{e^{(\nu-4\gamma)t}-1\over e^{\nu t}-1}.
\end{equation}
For the star phylogeny the genetic distance is always $t$ and
therefore $G_2^*(t)=e^{-4\gamma t}$. Here the relevant parameter is
the normalized mutation rate $\omega=\gamma/\nu$. Again, there exists
a critical point $\omega_c=1/4$. For smaller than critical mutation
rates, $\omega<\omega_c$, correlations due to the tree morphology are
not pronounced, $G_2(t)\propto G_2^*(t)$.  On the other hand, when
$\omega>\omega_c$, strong correlations are generated and $G_2(k)\sim
e^{-\nu t}$ is exponentially larger than $G_2^*(t)$. We conclude that
the behavior found for the deterministic case is robust.

\section{Multistate Sequences}

We now consider larger alphabets. Previously, the two states satisfied
$\sigma^2=1$. A natural generalization is to $\sigma^n=1$, i.e, the
$n$th order roots of unity $\sigma=e^{i2\pi l/n}$ with
$l=0,1,\ldots,n-1$. Previously, with probability $p$ the mutation
$\sigma\to \tau\sigma$ occurred with $\tau=e^{i\theta }$ and
$\theta=\pi$.  We thus impose the same transition but with
$\theta=2\pi/n$. This can be viewed as a clockwise rotation in the
complex plane by an angle $\theta$.  Since the states are now complex,
the definition of the pair correlation is now
\begin{equation}
\label{conjug}
G_2(k)=\langle\langle \bar\sigma_i\sigma_j\rangle\rangle, 
\end{equation}
with $\bar\sigma$ the complex conjugate of $\sigma$. The real part of
$\bar\sigma_i\sigma_j$ gives the inner product of the two-dimensional
vectors corresponding to $\sigma_i$ and $\sigma_j$, respectively. 

Consider the average $\langle \bar\sigma_3\sigma_4\rangle$ in Fig.~1.
Using the $\tau$ variables and $\bar\tau\tau=\bar\sigma\sigma=1$ one
has $\langle \bar\sigma_3\sigma_4\rangle=
\langle\bar\sigma_0\bar\tau_1\bar\tau_3\sigma_0\tau_1\tau_4\rangle=
\langle\bar\tau_3\tau_4\rangle=
\langle\bar\tau_3\rangle\langle\tau_3\rangle=
\overline{\langle\tau\rangle}\langle\tau\rangle=
|\langle\tau\rangle|^2$. All of our previous results hold if one
replaces the average $\langle\tau\rangle$ with its magnitude
\hbox{$\alpha=|\langle\tau\rangle|=|1-p(1-e^{i\theta})|=
\sqrt{1-2p(1-p)(1-\cos \theta)}$}. Furthermore, it is sensible to 
consider arbitrary phase shifts $0<\theta<2\pi$ since the identity
$\bar\sigma\sigma=\bar\tau\tau=1$ rather than $\sigma^n=\tau^n=1$ was 
used to evaluate  correlations.

The critical point is determined from the condition $\langle
r\rangle\alpha^2=1$.  This equation has a physical solution only when
$2\varphi<\theta<2(\pi-\varphi)$ with the shorthand notation 
\begin{equation} 
\varphi=\cos^{-1}\sqrt{1\over \langle r\rangle}.
\end{equation}
In terms of the number of states, this translates to 
\begin{equation} 
{\pi\over \pi-\varphi}<n<{\pi\over \varphi}.
\end{equation}
Hence, the transition may or may not exist depending on the details of
the model, in this particular ``clock'' model case, the number of
states. As we have seen before, correlations become less pronounced when
the number of ancestors increases. Indeed, the transition always
exists in the limit $\langle r\rangle\to 1$, while the transition is
eliminated in the other extreme $\langle r\rangle\to \infty$. When the
transition do occur, the following critical mutation probability is
found
\begin{equation}
\label{clock1}
p_c={1\over 2}\left[1-\sqrt{1-{\sin^2\varphi\over \sin^2{\theta\over 2}}}\,\,
\right].
\end{equation}
Indeed, Eq.~(\ref{pc1}) is reproduced in the two state case
($\theta=\pi$). This turns out to be the minimal critical point,
$p_c\geq(1-\sqrt{1/\langle r \rangle})/2$, reflecting the fact that
transition $\sigma\to -\sigma$ provides the most effective mutation
mechanism.

Interestingly the transition is restored when both the mutation and
the duplication processes occur continuously in time. In this
continuous description duplication occurs with rate $\nu$ and the
mutation \hbox{$\sigma\to e^{i\theta}\sigma$} occurs with rate
$\gamma$.  The autocorrelation \hbox{$A(t)=\langle \sigma(0)\bar
\sigma(t)\rangle=\exp\left[-\gamma(1-e^{-i\theta})t\right]$} is found
from its time evolution \hbox{$\dot
A(t)=-\gamma(1-e^{-i\theta})A(t)$}.  It can be easily shown from the
definition of the pair correlation (\ref{conjug}) that $A^2(\tau)$
should be replace with \hbox{$|A(\tau)|^2=\exp\left[-2\gamma(1-\cos
\theta)\tau\right]$} in the integral (\ref{integ}). Comparing with the
previous section results, we see that the effective mutation rate is
now $\gamma(1-\cos \theta)/2$. As a result, the location of the
critical point is increased by a factor $2/(1-\cos \theta)$. Using the
normalized mutation rate $\omega=\gamma/\nu$ one finds
\begin{equation}
\omega_c={1\over 2(1-\cos \theta)}.
\end{equation} 
This critical point increases with the number of states, and it
diverges according to $\omega_c\simeq (n/2\pi)^2$ when $n\to
\infty$. This behavior is intuitive as one expects that mutations
between a large number of states diminishes correlations and
consequently, phylogenetic effects.
             
\section{Two-site Correlations}

When sequences are not of unit length, {\it i.e.}, when there are two or more
sites per sequence, the results can be used to characterize a
correlation measure quantifying the interaction between sites. Assume
there are two or more sites per sequence and that the sites evolve
independently of each other. Denote the state of position $a$ in
sequence $i$ as $\sigma_i^a$, and similarly denote the state of
position $b$ in sequence $i$ as $\sigma_i^b$. If the sequences were
not related by a phylogenetic tree, but instead were independent
samples drawn from a given distribution, then the following quantity
defined on a finite set of $N=2^k$ samples specifies a two-site
correlation measure:
\begin{equation}
\label{rhodef}
\rho= {1\over N} \sum_i \sigma_i^a \sigma_i^b-{1\over N^2}\sum_i \sigma_i^a\sum_j \sigma_j^b. 
\end{equation}
Correlation between sites $a$ and $b$ is indicated by a non-zero value
of $\rho$.

The quantity $ \rho$ is well defined also when the sequences are
related by a phylogenetic tree. Due to the assumption of independent
positions, the mean of $\rho$ over all realizations vanishes $\langle
\rho\rangle=0$. This behavior is independent of the tree
morphology. To see the effects of the phylogeny, one needs to consider
fluctuations, {\it i.e.}, the variance $\Delta\rho=\langle\rho^2\rangle$  
\begin{eqnarray}
\Delta\rho&=&
\Bigl\langle\Big({1\over N}\sum_i \sigma_i^a \sigma_i^b\Big)^2\Bigr\rangle 
-\Bigl\langle\Big({1\over N}\sum_i \sigma_i^a\Big)^2
\Big({1\over N}\sum_i \sigma_i^b\Big)^2\Bigr\rangle\nonumber\\
&=&{1\over N^2}  \sum_{ij} \langle \sigma_i^a \sigma_j^a\rangle 
\langle \sigma_i^b\sigma_j^b \rangle 
-{1\over N^4}\sum_{ij}\langle\sigma_i^a \sigma_j^a\rangle
\sum_{kl}\langle\sigma_k^b \sigma_l^b\rangle\nonumber\\
&=&{1\over N^2}\Big[N+\sum_{i\ne j}\langle\tau\rangle^{2d_{i,j}}\Big]
-{1\over N^4}\Big[N+\sum_{i\ne j}\langle\tau\rangle^{d_{i,j}}\Big]^2\nonumber. 
\end{eqnarray}
The first equality in the above equation was obtained by rewriting
Eq.~(\ref{rhodef}) as $\rho=\rho_1-\rho_2$ and noting that $\langle
\rho_1\rho_2\rangle=\langle \rho_2^2\rangle$. The final expression can be
simplified using $\sum_{i\ne
j}\langle\tau\rangle^{2d_{i,j}}=N(N-1)G_2(\alpha^2,k)$ with
$G_2(\alpha^2,k)$ the pair correlation of Eq.~(\ref{g2ka}), considered
as a function of $\alpha^2$. The following expression for the
variance is obtained
\begin{equation}
\Delta\rho=\Big[{1\over N}+\big(1-{1\over N}\big)G_2(\alpha^2,k)\Big]
-\Big[{1\over N}+\big(1-{1\over N}\big)G_2(\alpha,k)\Big]^2.
\end{equation}

For the star morphology the leading order of the fluctuations is
independent of the mutation rate and it scales as the familiar
$N^{-1}$. For the binary tree morphology, there are again two regimes,
characterized by $p>p_c$ or $p<p_c$ where $p_c$ is now defined by
$2\alpha^4=1$, {\it i.e.},
\begin{equation}
\label{pc2}
p_c={1\over 2}\left( 1-{\left({1\over 2}\right)}^{1\over 4}\right).
\end{equation}
When $p<p_c$, the phylogeny plays a significant role and the variance
is exponentially enhanced $\Delta \rho\sim \alpha^{4k}$, while when
$p>p_c$, the variance is still statistical in nature $\Delta\rho\simeq
A\Delta^*\rho$ with $A>1$. Hence, it is more likely to observe large
values of $\rho$ in the tree morphology than it is in the star
morphology, even when the sites evolve independently. Since
correlations and variance play opposite roles, they are influenced in
different ways by the phylogeny.

\section{Summary}

In summary, we have studied the influence of the phylogeny on
correlations between the tree's nodes. In general, for sufficiently
small mutation rates, the morphology plays a minor role. For
sufficiently high mutation rates large correlations that can be
attributed to the phylogeny may occur. The transition between the two
regimes of behavior is sharp and is marked by a critical mutation
rate. Below this critical point all correlations are well described by
the average, while above it, correlations decay much slower than the 
average. Underlying this transition is the competition between the
multiplicity and the the degree of correlations between genetically
close and distant leafs. This competition also leads to larger
fluctuations in the correlation between different sites, even when
these evolve independently.

We have also seen that this behavior is robust and appears to
be independent of many details of the model. While the overall
behavior generally holds, specific details such as the location of the
critical point and the decay rate in the regime $p>p_c$ depend on a
specific tree dependent parameter: the average number of children.

The above results can be extended in several directions. It will be
interesting to see whether the recursive methods can be generalized to
stochastic tree morphologies and in particular to the continuous time
case. This methods should still be applicable even  when the mutation
rates are time dependent or disordered. In such cases it will be
interesting to determine which parameters determine the critical
point, the decay constants, etc. 

Correlations can serve as useful measure of the diversity of a system
since small correlations indicate large diversity and vice versa.  If
the diversity can be measured in an experiment where the phylogeny is
controlled, its time dependence can be used to infer the mutation
probability. Similarly, if the mutation probability can be controlled,
than the degree of correlation/diversity can be used to infer
characteristics of the phylogeny. Thus, our results may be useful for
inferring statistical properties of actual biological systems.

\bigskip
This research is supported by the Department of Energy under contract
W-7405-ENG-36.

\end{multicols} \end{document}